\documentclass[twoside]{article}

\usepackage{lipsum} 
\usepackage{aas_macros}

\usepackage[sc]{mathpazo} 
\usepackage[T1]{fontenc} 
\usepackage{microtype} 

\usepackage[hmarginratio=1:1,top=32mm,columnsep=20pt]{geometry} 
\usepackage{multicol} 
\usepackage[hang, small,labelfont=bf,up,textfont=it,up]{caption} 
\usepackage{booktabs} 
\usepackage{float} 
\usepackage{hyperref} 
\usepackage{natbib} 
\usepackage{graphicx}
\usepackage{caption} 
\usepackage{subcaption} 
\usepackage{lettrine} 
\usepackage{paralist} 
\usepackage{wasysym}
\usepackage{stmaryrd}

\usepackage{abstract} 

\usepackage{color}

\usepackage{ulem}

\usepackage{titlesec} 
\renewcommand\thesection{\Roman{section}} 
\renewcommand\thesubsection{\thesection.\arabic{subsection}} 
\titleformat{\section}[block]{\large\scshape\centering}{\thesection.}{1em}{} 
\titleformat{\subsection}[block]{\large}{\thesubsection.}{1em}{} 

\usepackage{fancyhdr} 
\usepackage{color}

\title{\vspace{-15mm}\fontsize{24pt}{10pt}\selectfont\textbf{UFOs: Just Hot Air or Something Meteor?}} 

\author{
\large
\textsc{Michael B. Lund$^1$}\\
\normalsize $^1$Caltech/IPAC-NExScI\\
\normalsize \href{mailto:editor@actaprimaaprilia.com}{editor@actaprimaaprilia.com} 
\vspace{-5mm}
}
\date{}

\begin{document}

\maketitle 

\thispagestyle{fancy} 


\begin{abstract}

\noindent For much of February 2023, the world was in panic as repeated balloon-like unidentified flying objects (UFOs) were reported over numerous countries by governments that often responded with military action. As a result, most of these craft either escaped or were destroyed, making any further observation of them nearly impossible. These were not the first time balloon-like objects have loomed over Earth, nor are they likely to be the last. This has prompted us to push for a better understanding of UFOs. First we demonstrate that the distribution of balloon incidents and other UFO reports are consistent with being drawn from the same geographic distribution, and further that both of these distributions are consistent with the areas of the Earth that feature the jet stream. Second we show that there are more UFO sightings during meteor showers, as we would expect if meteor showers, already a known source of extraterrestrial material, are being used to provide some manner of distraction to help alien craft enter the Earth's atmosphere without drawing undue attention. These links between alleged balloon incidents, UFO reports, and meteor showers establish a transport pipeline for alien craft from interplanetary and possibly interstellar space to the Earth's surface.

\end{abstract}


\begin{multicols}{2} 

\section{Introduction}
\lettrine[nindent=0em,lines=3]{T}he history of humans and documented observations of UFOs arguably goes back thousands of years \citep{vonDaniken1969}. With such a broad topic, ranging from historical artifacts pointing to their presence in the past \citep{Clarke1968, Kubrick1968} to current close encounters with UFOs \citep{Hynek1972, Spielberg1977}, we choose to focus on specifically exploring the connections between UFO sightings and reports of lighter-than-air objects and understanding how these sightings can be related to more well-characterized extraterrestrial objects.

There's a long history of objects that have straddled the UFO and lighter-than-air categories; the most famous of which is the Rosewell incident which was initially reported as a flying saucer that had been captured by the Roswell Army Air Field \citep{RoswellDailyRecord1947}. However, government officials would quickly reverse course and claim that the object in question was actually a weather balloon \citep{FortWorthStarTelegram1947}.

This was far from the first time that UFOs have been linked to lighter-than-air craft, however, as mysterious airships had been reported as parts of larger waves of UFOs on several occasions. In 1896 and 1897, large regions of the United States saw numerous reports of unidentified craft, often with descriptions matching blimps \citep{Bartholomew1998, Kocis2020} These reports began in California but quickly spread nationwide \citep{Reece2007}.

It is to be expected, then, that any such craft would be impacted significantly by wind, and there are key benefits to traveling along the jet stream as this is both faster and more energy efficient \citep{Hunt2019}. The jet stream can be roughly defined as the regions between the subtropical jet at $30^{\circ}$ N/S and the polar jet at $50-60^{\circ}$ N/S \citep{NOAA2022}. The entire range from $30^{\circ}$ to $60^{\circ}$ is consistent with the area "categorized by velocity maxima as belonging to a subtropical or polar jet" \citep{PenaOrtiz2013}. This gives us, then a terrestrial region where we would expect most UFO sightings to occur if they are, indeed, still broadly representative of a population of lighter-than-air craft.

Now that we have an understanding of how terrestrial locations could correlate with UFO sightings, we must also understand where outside Earth we could expect to be associated with UFOs. Of course, there are some bodies in the Solar System that have been proposed to be alien craft in their own right, such as Phobos \citep{Shklovsky1962, Houston1959}, 1I/`Oumuamua \citep{Loeb2022}, and asteroid 31/439 \citep{Clarke1973}. These are not objects, however, that could be confused for objects entering the Earth's atmosphere. Instead, we are more interested in situations where an alien craft could be shielded on a planetary approach by being close to, behind, on, or inside of a naturally occurring body. Two prominent candidates that have been put forward for this are asteroids \citep{Kershner1980} and comets \citep{Muren1987, Achenbach1997}. These are particularly well-suited because they provide cover via meteor showers.

Meteor showers have been documented going back hundreds of years, at least as far back as August 1583, when Songhay
historian, Mahmud Al Kati recorded what appears to be a meteor shower occurring after midnight (from an interview in \citet{delCastillo2002}):
\begin{quote}
In the year 991 [1583 A.D.] in God’s month of Ragab the Goodly [August] after half the night passed, stars flew around the sky as if fire had been kindled in the whole sky -- east, west, north and south. It became a mighty flame lighting up the earth and people were extremely disturbed by that. It continued until after dawn.
\end{quote}
However, the first to make the strong argument that meteor showers were necessarily of extraterrestrial origins may not have been prior to the 1830s, when Denison Olmsted synthesized multiple observations on November 13, 1833 of what is now known as the Leonids \citep{Olmsted1834}. If UFOs are indeed needing to have some sort of distraction in order to make their approaches to Earth, we should then expect that there are more UFO sightings associated with meteor showers than the baseline rate of sightings.

In this paper, we look at over 80,000 UFO sightings and how these correlate with 2023 reports of high-altitude balloons and the jet stream in order to examine if UFOs and alleged high-altitude balloons may, in fact, represent the same underlying population of craft. We also try to better understand the origins of these objects by looking at how these observations relate to meteor showers. In Section~\ref{Data}, we provide overviews of the various data sets that we use for UFO sightings, balloon incidents, and meteor showers. We then compare these data sets in Section~\ref{Analysis} and discuss the implications of this in Section~\ref{Discussion}.
Finally we summarize this work in Section~\ref{Summary}.

\section{Methods} \label{Methods}

\subsection{Data} \label{Data}

\subsubsection{UFO Sightings}
\begin{figure*}[!htb]
  \begin{center}
   \includegraphics[width=0.95\textwidth]{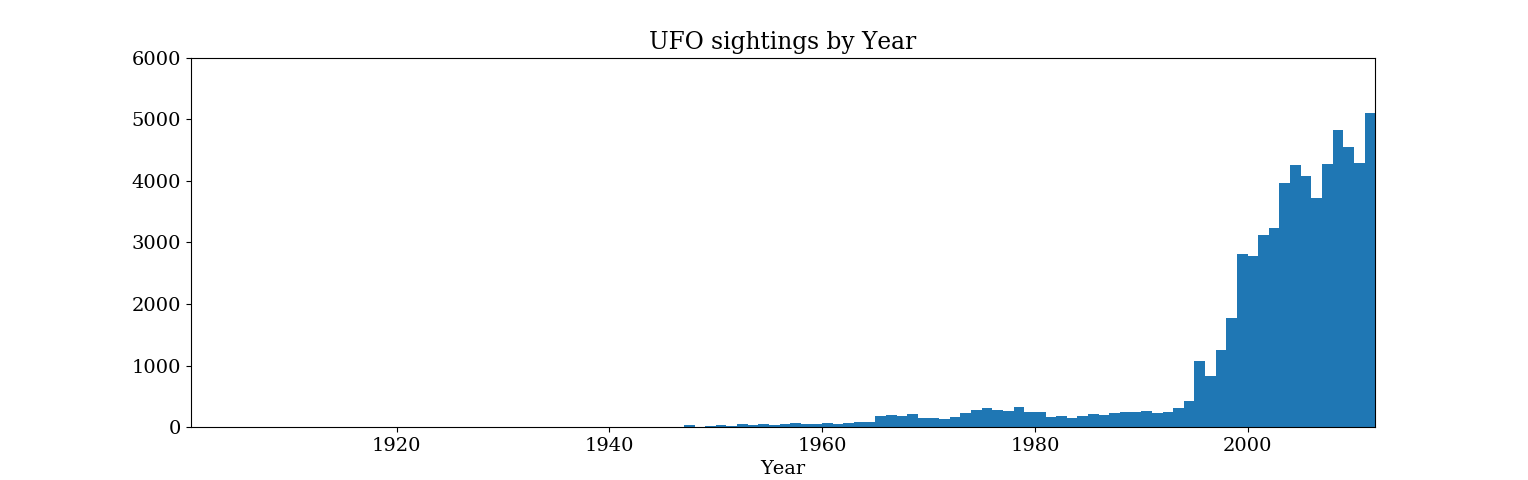}
  \end{center}
  \caption{UFO sightings as a function of year in NUFORC data set.}
  \label{fig:UFO4}
\end{figure*}
\begin{figure*}[!htb]
  \begin{center}
   \includegraphics[width=0.95\textwidth]{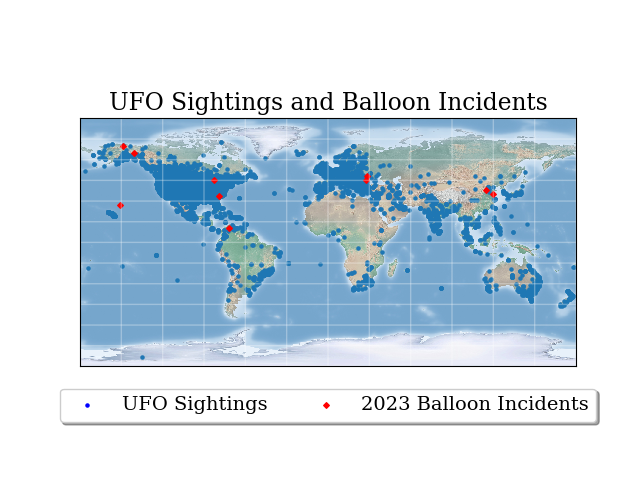}
  \end{center}
  \caption{UFO Sightings in the NUFORC data set are marked with blue dots. 2023 "Balloon" incidents are marked with red diamonds.}
  \label{fig:UFO1}
\end{figure*}

There have been multiple attempts to document the scope of UFO sightings, going back as far as the Condon Report's summary of the United States Air Force's Project Blue Book database of UFO sightings \citep{Condon1969}. More recent attempts include the over 80,000 sightings that have been extracted from the  National UFO Reporting Center (NUFORC) database and made available as a kaggle data set \footnote{\url{https://www.kaggle.com/datasets/NUFORC/ufo-sightings/versions/2}}. In this data set, sightings include a description of the sighting as well as the date and time of the sighting and the location, including latitude and longitude. The dates are subsequently converted to day of year as part of our analysis. This data set has a very long baseline, with the earliest reported sightings from 1906, but the vast majority of sightings come from the last decade or so of the data set, as shown in Figure~\ref{fig:UFO4}. The full geographic distribution of UFO sightings on Earth are displayed in Figure~\ref{fig:UFO1}. This data set also includes parameters such as shape and duration, however these are beyond the cursory scope of this paper.

\subsubsection{2023 "Balloon" Incidents}
Over the course of February 2023, there were numerous reports of high-altitude objects over multiple countries and creating significant concerns about airspace security\footnote{\url{https://en.wikipedia.org/wiki/List_of_high-altitude_object_events_in_2023}}. According to most government claims, the explanation for these events were alleged to be balloons of some kind, but yet the true nature of these objects often were unidentified, either because these objects were destroyed or because these objects escaped. We include a full table of these objects in Table~\ref{tab1:balloons}, using locations and coordinates corresponding to the last position reported.

\begin{table*}[htb]
\caption{February 2023 Balloon Incidents}
\label{tab1:balloons}
\begin{center}
\begin{tabular}{|llrr|}
\hline
Date & Location  & Latitude & Longitude \\ \hline
02/03/23 & Maracaibo, Venezuela & 11 & -72 \\
02/04/23 & South Carolina, USA & 34 & -79 \\
02/10/23 & Deadhorse, Alaska, USA & 70 & -149 \\
02/11/23 & Yukon, Canada & 65 & -141 \\
02/12/23 & Lake Huron, Canada & 45 & -82 \\
02/12/23 & Shandong, China & 35 & 120 \\
02/14/23 & Romania & 45 & 28 \\
02/14/23 & Moldova & 48 & 28 \\
02/16/23 & Shijiazhuang, China & 38 & 115 \\
02/16/23 & Hawaii, USA & 27 & -151 \\
\hline
\end{tabular}
\end{center}
\end{table*}

\subsubsection{Meteor Showers}
While there are nearly three dozen meteor showers, we only look at prominent meteor showers as defined by an article in Sky \& Telescope forecasting 2022 meteor showers and partially reproduced here in Table~\ref{tab1:showers} \citep{Beatty2021}. From this list of meteor showers, we take the peak or central date of each meteor shower and then convert this to a corresponding day of year. We also include the listed host bodies; of the twelve major meteor showers, 10 of the identified objects are unambiguously identified as comets and the other two are asteroids (denoted with italics on the table). Of the two remaining objects, 2003 $EH_{1}$ is a near-Earth object that has also been identified as potentially being the extinct remnants of comet C/1490 Y1 and linking the Quadrantids with a comet as well \citep{Jenniskens2004}.
\begin{table*}[htb]
\caption{Meteor Showers and Associated Bodies}
\label{tab1:showers}
\begin{center}
\begin{tabular}{|llll|}
\hline
Meteor Shower  & Central Date & Day of Year & Parent Body \\ \hline
Quadrantids & Jan 3 & 3 & \textit{2003 $EH_{1}$}\\
Lyrids & Apr 22 & 112 & Thatcher (1861 I)\\
Eta Aquariids & May 6 & 126 & 1P/Halley \\
Tau Herculids & May 31 & 151 & 73P/Schwassman-Wachmann \\
Delta Aquariids & Jul 30 & 211 & 96P/Machholz \\
Perseids & Aug 13 & 225 & 109P/Swift‑Tuttle \\
Southern Taurids & Oct 13 & 225 & 2P/Encke \\
Orionids & Oct 21 & 294 & 1P/Halley \\
Northern Taurids & Nov 5 & 309 & 2P/Encke \\
Leonids & Nov 18 & 322 & 55P/Tempel‑Tuttle \\
Geminids & Dec 14 & 348 & \textit{3200 Phaethon} \\
Ursids & Dec 22 & 356 & 8P/Tuttle \\
\hline
\end{tabular}
\end{center}
\end{table*}

\subsection{Analysis} \label{Analysis}

We begin by first exploring the overlap of UFO sightings and balloon incidents, as well as how this corresponds to the earth's jet streams, as the jet streams would be the areas where balloons, or balloon-like objects, would be most useful. We begin by looking at what fraction of the Earth is represented by the band of the atmosphere where the jet streams can be found.

\begin{figure*}[!htb]
  \begin{center}
   \includegraphics[width=0.95\textwidth]{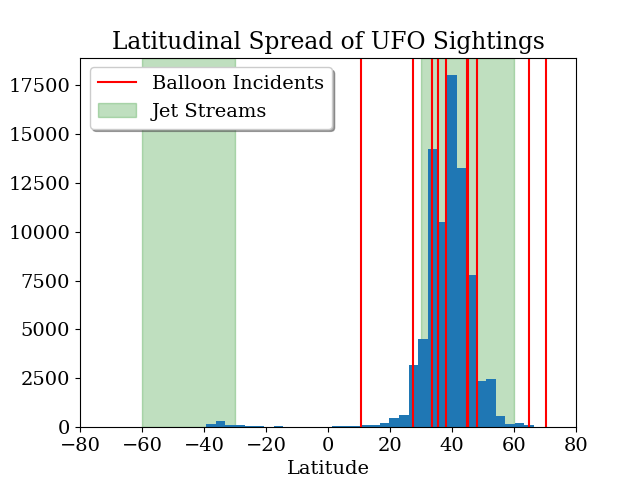}
  \end{center}
  \caption{A histogram showing all UFO sightings binned by latitude in blue. Additionally, the standard latitude range of the jet stream (30-60 degrees N/S) are represented by the green regions. Alleged balloon incidents are marked in red and using the locations in Table~\ref{tab1:balloons}.}
  \label{fig:UFO3}
\end{figure*}

Per Stack Exchange\footnote{User Rahul Madhavan on April 21, 2021 in post \url{https://math.stackexchange.com/questions/4102850/area-of-surface-between-two-lines-of-latitude}} (as is the standard way to resolve most moderate math questions), the area of a sphere between two latitudes can be expressed with the following equation:
\begin{equation}
    Area = 2 \pi r \int_{\theta=lat_1}^{lat_2} \!  cos (\theta) d(\theta)
\end{equation}
This can then be simplified to:
\begin{equation}
    Area = 2 \pi r (sin(lat_1) - sin(lat_2))
\end{equation}
However, as we only are interested in what fraction of the Earth's surface this corresponds to, this is more usefully expressed with the following:
\begin{equation}
    Fractional Area = \frac{(sin(60^{\circ}) - sin(30^{\circ}))}{(sin(90^{\circ}) - sin(0^{\circ}))}
\end{equation}
When this is calculated, we get that 36.6\% of the Earth's surface is in the regions that would have the jet streams overhead. In contrast, though, we find that 91.1\% of UFO sightings took place in this same region, and this heavy overlap is visible in Figure~\ref{fig:UFO3}.
This represents a very high correlation between the location of the jet stream and UFO sightings.

In Figure~\ref{fig:UFO3} we can also see that not only is there significant overlap between the UFO sightings distribution (in blue) and the jet streams (green regions), but also that the "balloon" incidents in 2023 largely took place in the same latitudes. Six of the ten balloon incidents occurred within the latitude range of the jet streams.

We next directly compared the "balloon" incidents and the UFO sightings. We conducted a Kolmogorov-Smirnov test to see if the "balloon incidents could represent the same latitude distribution as the UFO sightings \citep{Kolmogorov1933, Smirnov1948}. Given the p-value of 0.16, we are unable to reject the null hypothesis that UFO sightings and "balloon" incidents are the same underlying distribution and so it is still consistent that these represent different kinds of observations of the same population of objects.

\begin{figure*}[!htb]
  \begin{center}
   \includegraphics[width=0.95\textwidth]{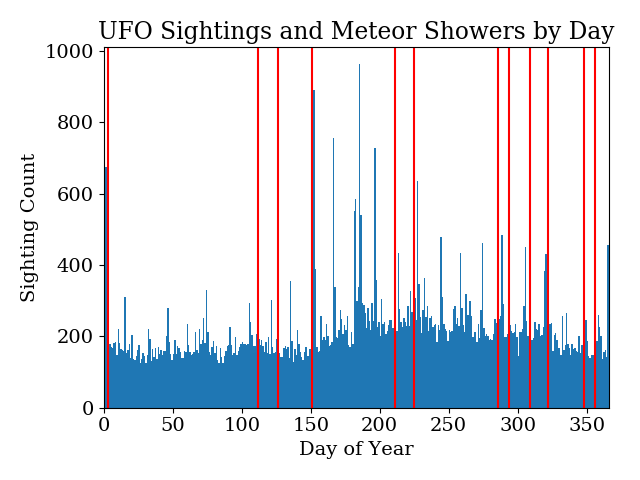}
  \end{center}
  \caption{UFO sightings as a function of day of year represented in blue. Red vertical lines correspond to the meteor showers that are included in Table~\ref{tab1:showers}.}
  \label{fig:UFO2}
\end{figure*}

Now that we have statistically linked the "balloon" incidents, UFO sightings, and the jet stream, we can then move on to linking these terrestrial observations with their extraterrestrial sources. To do this, we look at how UFO sightings correspond with meteor showers as in Figure~\ref{fig:UFO2}.
The median number of UFOs observed on a single day of the year (over the cumulative data set, not per year) was 198 reports. For the meteor showers in Table~\ref{tab1:showers}, we look at the number of UFOs reported during each meteor shower by first averaging over three days for each shower (the central date $\pm$ 1 day) and then looking at the median of all meteor showers. We find that during meteor showers, the median number of UFO reports per day of the year (again, a cumulative value) is 255.5 reported sightings, or a startling 29\% increase over the median number across all days of the year. This provides a powerful link between UFO sightings and meteor showers.

\section{Discussion}\label{Discussion}
Establishing the link between UFOs and modern incidents that have been reported as high-altitude balloons provides a key insight into UFOs. Most importantly, this explains why we observe UFOs so frequently between $30^{\circ}$ and $60^{\circ}$ N/S, where trade winds would most benefit travel of lighter-than-air craft that can be carried by the wind.

Understanding the relationship between UFOs and meteor showers provide us with an additional tool, as we can use bulk studies of meteors to try to also provide incidental data of alien lighter-than-air craft. One of astronomy's most valuable tools in understanding objects, spectroscopy, can be brought to bear on UFOs as meteor spectra have been recorded for well over a century \citep{Pickering1897}, and these sort of observations have been well-suited to studies of composition \citep{Millman1933}. This early characterization was used to identify metals in these meteors, but hydrogen and helium were notable non-detections given their common use in terrestrial craft \citep{Millman1937}.

In more recent observations, hydrogen has been detected in cometary meteroids, but it has not been detected in asteroidal meteoroids, which suggests that detections of hydrogen represent tracers of water and organic materials \citep{Matlovic2022}. All indications suggest that these meteorites only contain trace amounts of hydrogen, however, and that the bulk composition of hydrogen is only between 0.5\% and 1.5\% of the total mass; this hydrogen content is far below what would be needed for there to be enough hydrogen to actually provide lighter-than-air capability in any craft \citep{Lee2021}.

With lighter-than-air gases hydrogen and helium eliminated as options, the only option for these balloons to function being that these balloons must be heating some fashion of ideal gases in order to attain sufficient buoyancy for travel in the Earth's atmosphere \citep{Clapeyron1834}. The presence of these heated gases may suggest that infrared detection would be a possible approach for more effective detection and characterization of UFOs.

\section{Summary}\label{Summary}
In this work, we have established that both UFO reports broadly and the specific objects observed in February 2023 and sometimes described as high-altitude balloons appear to come from the same distribution of objects that are most likely to be located in parts of the Earth that feature the jet stream, where balloons would be most useful. We also demonstrate a link between the increase of meteors (during meteor showers) and the increase in reported UFOs. Combined, these results suggest that reported UFOs are frequently lighter-than-air craft that use meteor showers as a screen to enter the Earth's atmosphere and possibly use their host bodies to cloak their approaches towards Earth.

\section{Acknowledgements}
The author wishes to thank.Kyle Conroy for feedback on this manuscript.


This research has made use of NASA’s Astrophysics Data System.

Software: matplotlib \citep{Hunter2007}, numpy \citep{Oliphant2006}, pandas \citep{Mckinney2011}, scipy \citep{Virtanen2020}


\bibliographystyle{apalike}
\bibliography{main}


\end{multicols}

\end{document}